\documentclass[twocolumn,prb,showpacs]{revtex4}

\usepackage{graphicx}
\usepackage{rotating}
\usepackage{amsmath}
\usepackage{amsfonts}
\usepackage{amssymb}
\usepackage{enumerate}
\usepackage{longtable}
\usepackage{dcolumn}% Align table columns on decimal point
\usepackage{bbm}

\begin{document}

\title{Efficient Coherent Control by Sequences of Pulses of Finite Duration}

\author{G\"otz S. Uhrig}
\email{goetz.uhrig@tu-dortmund.de}
\affiliation{Lehrstuhl f\"{u}r Theoretische Physik I,
Technische  Universit\"{a}t Dortmund,
 Otto-Hahn Stra\ss{}e 4, 44221 Dortmund, Germany}

\author{Stefano Pasini}
\email{pasini@fkt.physik.tu-dortmund.de}
\affiliation{Lehrstuhl f\"{u}r Theoretische Physik I,
Technische  Universit\"{a}t Dortmund,
 Otto-Hahn Stra\ss{}e 4, 44221 Dortmund, Germany}

\date{\rm\today}

\begin{abstract}
Reliable long-time storage of arbitrary quantum states is a key
element for quantum information processing.
In order to dynamically decouple a spin or quantum bit from a dephasing
environment by non-instantaneous pulses, we introduce an optimized sequence of
$N$ control  $\pi$ pulses which are realistic in the sense
that they have a finite duration and a finite amplitude. We show that  
optimized dynamical decoupling is still applicable and 
that higher-order decoupling can be reached 
if shaped pulses are implemented. 
The sequence suppresses decoherence up to the order 
${\cal O}(T^{N+1})+ {\cal O}(\tau_\mathrm{mx}^M)$, 
with $T$ the total duration of the sequence and  
$\tau_\mathrm{mx}$ the maximum length of the pulses. The exponent 
$M\in\mathbb{N}$ depends on the shape of the pulse.
Based on existing experiments, a concrete setup for the 
verification of the properties of the advocated sequence is proposed.
\end{abstract}

\pacs{03.67.Pp, 82.56.Jn, 76.60.Lz, 03.65.Yz}

%% PACSes: to be redone - see also NMR
%% I (GSU) find the PACSes OK
% 03.67.Pp Quantum error correction and other methods for protection against 
% decoherence (see also 03.65.Yz Decoherence; open systems; quantum statistical
%  methods; for decoherence in Bose-Einstein condensates, see 03.75.Gg) 
% 03.67.Lx Quantum computation
% 03.65.Yz Decoherence; open systems; quantum statistical methods 
% (see also 03.67.Pp in quantum information; for decoherence in Bose-Einstein 
% condensates, see 03.75.Gg) 
% 03.65.Vf Phases: geometric; dynamic or topological  
% 76.60.Lz Spin echoes
% 82.56.Dj High resolution NMR
% 82.56.Jn Pulse sequences in NMR
% 33.25.k Nuclear resonance and relaxation: Atomic and Molecular Physics

\maketitle

\section{Introduction}

In  quantum information processing (QIP) and in nuclear magnetic
resonance (NMR) it is essential to be able to decouple
the quantum bit or the spin, respectively, from its environment. 
Both fields of research are of wide-spread interest and relevance.
In the former the ultimate goal is to realize reliable long-time storage
of quantum information with as low as possible error rates. This
is a prerequisite for QIP \cite{niels00,zolle05}. In the latter,
the high-precision measurement of nuclear spin dynamics is a long-standing
goal \cite{haebe76,freem98}.

Besides choosing well-isolated systems the application of
appropriately tailored sequences of control pulses 
\cite{haebe76,viola98,ban98}, i.e., dynamic
decoupling (DD), is one of the promising routes to this goal.
The basic idea goes back to Hahn's spin echo pulse which averages
 a static perturbance to zero \cite{hahn50}. 
For a dynamic environment, or bath,
sequences of pulses are required \cite{carr54,meibo58,viola98,ban98,viola99a}.
The early suggestions are essentially periodic in time.

Recently, the additional advantages of sequences with non-equidistant pulses
were discovered. Concatenation (CDD) can suppress unwanted couplings in a high 
power $T^l$ of the length of the sequence \cite{khodj05}. 
But for the method used in Ref.\ \onlinecite{khodj05},
the required number $N$ of pulses grows exponentially with $4^\ell$. 
For pure dephasing, it was shown that
this growth can be reduced decisively to a linear one $l\propto N$
if the instants of $\pi$ pulses were chosen according to
\begin{equation}
\label{eq:switch-t-UDD}
t_j = T \sin^2(j\pi/(2N+2)),
\end{equation}
which is called UDD (Uhrig DD).
The relation \eqref{eq:switch-t-UDD} was derived  for a spin-boson model 
\cite{uhrig07} where it was observed that no details of the model entered. 
On the basis of numerical evidence and finite order recursion
it was conjectured  that \eqref{eq:switch-t-UDD} is applicable
to any dephasing model \cite{lee08a,uhrig08}. This claim was
finally proven\cite{yang08} for any order in $T$. For various simulated
classical noise
spectra the experimental verification of the theoretical results  was achieved 
 \cite{bierc09a,bierc09b,uys09b} by microwave
control of the transition in Be ions. 
{The suppression of the decoherence of the electron spin
of hydrogen radicals was investigated by electron spin resonance in crystals 
of irradiated malonic acid \cite{du09}. The decoherence was due to the
quantum noise induced by nuclear spins. Again, the UDD proved superior
to standard sequences.}

For general decoherence, concatenation of
the UDD sequence (CUDD) can be used \cite{uhrig09b}.
For a suppression of the decoherence up to $T^\ell$, 
the number of pulses grows as  $2^\ell$
which is an improvement by a square root with respect to the CDD of 
Ref.\ \onlinecite{khodj05}. A more efficient
scheme, called quadratic DD (QDD), 
which requires only a quadratically growing number of
pulses, has been proposed very recently based on numerical 
\cite{west09} and analytical evidence \cite{pasin09c}.

All these  sequences (periodic DD  \cite{viola99a},
CDD \cite{khodj05}, UDD \cite{uhrig07}, CUDD \cite{uhrig09b},
and QDD \cite{west09,pasin09c})
 rely on instantaneous, thus idealized,
pulses. This problem was realized early on and ongoing
research investigates pulses
of finite duration $\tau_\mathrm{p}$ 
\cite{sengu05,motto06,pasin08a,pasin08b,karba08} 
and sequences of such pulses \cite{viola03,khodj07,pryad08a,pryad08b,santo08}.
Eulerian DD  \cite{viola03,khodj09a} is designed to annihilate the first
order of a Magnus expansion over the whole sequence. Thus corrections
of the order of $T\tau_\mathrm{p}$ are not excluded. Similar caveats apply
to many other sequences  \cite{khodj07,pryad08a,pryad08b,santo08}.
Also the experimental realizations in Refs.\ 
\onlinecite{bierc09a,bierc09b,uys09b} have to take into account
that real pulses cannot be instantaneous because
the control amplitudes are necessarily bounded.

Our aim here is to derive an optimized
sequence with UDD properties which relies on realistic 
pulses of finite duration and which is adapted to these real pulses.
We do not provide a general scheme to use
pulses of bounded control for arbitrary DD sequences.
If the shape is appropriately designed, 
the pulse can be approximated as an instantaneous one up to 
${\cal O}(\tau_{\mathrm p}^M)$. For $M=3$ explicit results  
were derived in Ref.\ \onlinecite{pasin09a} while a recursive
scheme for arbitrary $M$ has been proposed recently \cite{khodj09c}.
As far as the correction ${\cal O}(\tau_{\mathrm p}^M)$ is negligible, 
the proposed sequence displays the same exact analytic properties as the 
UDD sequence of ideal, instantaneous pulses.

We approach the problem hierarchically. That means that we use the pulses and 
the periods of free evolution as building blocks for the sequence.
First, the properties of the pulses are derived and discussed.
Second, these properties are used in the sequence.
To this end, we exploit scaling in two
independent variables, namely the durations $\tau_\textrm{p}$ of the pulses,
whose maximum is $\tau_\textrm{mx}$, and the total duration of the sequence
denoted by $T$. Note that these two time scales
are largely independent, both in theory and in experiment,
because the pulses are not applied back to back.
The only constraint is that $T$ must be larger than the sum
of the pulse durations $\tau_{\textrm{p},j}$ (belonging to pulse $j$)
\begin{equation}
\label{eq:constraint}
T\ge \sum_{j=1}^N \tau_{\textrm{p},j}.
\end{equation}
 Relying on pulses, which cancel all orders $m<M$, the whole sequence avoids
all mixed terms  $T^{n}\tau_\mathrm{mx}^m$, where
$n\le N+1$ and $m<M$. 
Hence  important progress over existing proposals
\cite{viola03,khodj07,pryad08a,pryad08b,santo08} is achieved.

\section{Model} 

We start from the Hamiltonian
\begin{equation}
\label{eq:hamilton}
H=\hat A_0 + \sigma_z \hat A_1 \widetilde F(t)
\end{equation}
where $\sigma_z$ is the $z$ component of the Pauli matrices. It is acting
on the $S=1/2$ spin or, generally, on the two-level system which 
represents a qubit. The operators $\hat A_i$ act on the bath
only; they may also be c-numbers. 
We consider any kind of bath with bounded operators for the
sake of the mathematical argument $||H||\leq \gamma<\infty$,
where $||\cdot ||$ is any appropriate operator norm
which remains invariant under unitary transforms. 
We expect that the order of suppression of the decoherence holds
for any bath which can be approximated  by bounded baths,
i.e., the bath should have a hard high-energy cutoff.

No spin flip terms are included in \eqref{eq:hamilton} implying
an infinite spin-lattice relaxation time
$T_1$. It is an excellent approximation if $T_1\gg T_2$  where
$T_2$ is the dephasing time. Such a situation is achieved
in the rotating reference frame of a system where 
the two levels with eigenvalues $\pm1$ of $\sigma_z$  lie energetically 
far apart. Longitudinal relaxation and general decoherence
will be addressed below.

Moreover, \eqref{eq:hamilton} is the effective Hamiltonian
in the interaction picture of the short control pulses
\cite{uhrig07,lee08a,uhrig08,yang08}.
Thus the switching function  $\widetilde F(t)\in\mathbb{R}$ appears.
The simplest example is an instantaneous $\pi$ pulse at $t=t_j$
which realizes a rotation about an axis perpendicular to
$\sigma_z$. Then $\widetilde F(t)$ changes its sign at $t=t_j$ abruptly
while it is constant elsewhere. A sequence of such pulses at the 
instants $\{t_j\}$ with $j\in\{1,2,\ldots,N\}$  implies 
$\widetilde F(t)=(-1)^j$ for $t\in\{t_j,t_{j+1}\}$ where we define $t_0=0$
and $t_{N+1}=T$.

{Our derivation is based on the bounded quantum model 
\eqref{eq:hamilton}. Thereby, classical Gaussian noise is treated to the
extent that it can be approximated by the quantum model \cite{uhrig08}.
Certainly, non-Gaussian classical noise, see for instance
Ref.\ \onlinecite{cywin08}, should be considered separately
which is beyond the scope of the present article.}

\section{Derivation}
\label{sect:dephasing}

Optimization of the sequence means to ask the question which
switching instants $t_j$ make the sequence $\{t_j\}$ most efficient.
For ideal instantaneous pulses, it was shown that the UDD instants  
\eqref{eq:switch-t-UDD} are optimum in the sense that the time evolution 
depends on the spin weakly \cite{uhrig07,lee08a,uhrig08,yang08}. 
The time evolution operator 
\begin{equation}\label{eq:defUpm} \hat U_\pm=\prod_{j=0}^N
e^{-i[\hat {A}_0\pm\hat {A}_1\tilde F(t_j)](t_{j+1}-t_j)}
\end{equation} 
for the eigenstates of $\sigma_z$
with eigenvalues $\pm 1$
depends on the spin only in a high power of $T$
\begin{equation}
\label{eq:prop-UDD}
\hat U_+ - \hat U_- = {\cal O}((\gamma T)^{N+1}).
\end{equation} 
The analytical derivation of \eqref{eq:prop-UDD}
is achieved by direct  time-dependent perturbation theory  (TDPT) \cite{yang08}
in powers of $t H$. Thus, if the $N+1$st power does not vanish it
is of order $(\gamma T)^{N+1}$.
The iterated time integrations of TDPT  
are conveniently expressed by the substitution 
$t:=T\sin^2(\vartheta/2)$ as integrations over the variable $\vartheta$.
The instants \eqref{eq:switch-t-UDD} are equidistant if
expressed in  $\vartheta$ because for $F(\vartheta):=
\widetilde F(T \sin^2(\vartheta/2))$ we have
\begin{equation}
\label{eq:switch-theta-UDD}
F_\mathrm{UDD}(\vartheta) = (-1)^j 
\quad  \mathrm{for} \quad \vartheta \in
\left(\frac{j\pi}{N+1},\frac{(j+1)\pi}{N+1} \right)
\end{equation}
with $j\in\{0,N\}$. Allowing $j$ to take all integer values
$j\in \mathbbm{Z}$ the function $F_\mathrm{UDD}$ becomes an odd
function with $F_\mathrm{UDD}(\vartheta+\pi/(N+1))=
-F_\mathrm{UDD}(\vartheta)$. 
Hence the Fourier series of $F_\mathrm{UDD}(\vartheta)$ comprises only odd 
$\sin$ harmonics $\sin(l(N+1)\vartheta)$ with $l\in\{1,3,\ldots\}$. The 
coefficients are $4/(\pi l)$. 
From this property, \eqref{eq:prop-UDD} is derived by
exploiting trigonometric addition theorems recursively \cite{yang08}.

The power of the UDD sequence has been demonstrated experimentally 
\cite{bierc09a,bierc09b}.
The noise, i.e., the coupling to the bath $\hat A_1$, is simulated,
so that it can be switched off during the pulse. Thereby, a partial solution
of the  finiteness of the pulse amplitudes is achieved.
But generally decoherence  processes  cannot be switched off. 
Using pulses of duration $\tau_\mathrm{p}$ with 
constant amplitude instead of instantaneous pulses introduces 
 an unwanted term of the order $\gamma\tau_\mathrm{p}$ at each rotation, i.e.,
linear in the pulse length. For a sequence of length $N$ these
corrections can accumulate to $N\gamma\tau_\mathrm{p}$ unless the
contributions of subsequent pulses cancel each other.

An improvement by one order in $\gamma\tau_\mathrm{p}$ is achieved by
the ersatz $\pi$ pulse  which makes the linear 
correction vanish. Then the time evolution operator $\hat U_\mathrm{p}$ 
for a pulse reads
\begin{subequations}
\label{eq:ansatz-dd}
\begin{eqnarray}
\label{eq:ansatz1}
\hat U_\mathrm{p}(t+\tau_\mathrm{p},t) &=&
\hat U_\mathrm{p}^\mathrm{ideal}(t+\tau_\mathrm{p},t)  + 
{\cal O}((\gamma\tau_\mathrm{p})^M)\qquad
\\
\label{eq:ideal}
\hat U_\mathrm{p}^\mathrm{ideal}(t+\tau_\mathrm{p},t) &=& 
e^{-\mathrm{i}(\tau_\mathrm{p}-\tau_\mathrm{s})H}
\hat{P}_\theta e^{-\mathrm{i}\tau_\mathrm{s}H}, 
\end{eqnarray}
\end{subequations}
where $M=2$. It is understood that $t$ marks the beginning of the
pulse and $t+\tau_\mathrm{p}$ its end.
It is important that
$H$ is the Hamiltonian of the total system, i.e., spin, bath,
and their mutual coupling, $\hat{P}_\theta$ is the ideal pulse
with $\theta=\pi$, and 
$\tau_\mathrm{s}$ is the instant when the approximated ideal pulse
occurs. In a sequence $\{t_j\}$, the instant
$\tau_\mathrm{s}$ is to be identified with the switching instants $t_j$.
No adjustment of the sequence takes place.
Relation \eqref{eq:ansatz-dd} can be achieved by shaping the pulse
appropriately \cite{pasin08a,pasin08b,karba08}.  Hence we can
set up a UDD sequence with more realistic pulses of the kind
\eqref{eq:ansatz-dd} for which the  deviations read
\begin{equation}
\label{eq:bound1}
\hat U_+^\mathrm{UDD} - \hat U_-^\mathrm{UDD} =
{\cal O}\left((\gamma T)^{N+1}\right) +
{\cal O}\left(N(\gamma \tau_\mathrm{mx})^M\right)
\end{equation}
with $M=2$.
The additivity of the corrections is a straightforward 
property of the unitary  evolution operators. If we
denote the UDD sequence made from the ideal pulses 
$\hat U_\mathrm{p}^\mathrm{ideal}(t+\tau_\mathrm{p},t)$
in \eqref{eq:ideal} 
by $\hat U_\pm^\mathrm{UDD,ideal}$, we know from \eqref{eq:ansatz1}
\begin{equation}
\label{eq:add1}
\hat U_\pm^\mathrm{UDD} = \hat U_\pm^\mathrm{UDD,ideal} +
{\cal O}(N(\gamma\tau_\textrm{mx})^M)
\end{equation}
for $N$ pulses. The unitary invariance of the norm $\gamma$
is used for each pulse. Next, we know from the properties of
the UDD sequence \cite{lee08a,uhrig08,yang08}
\begin{equation}
\label{eq:udd-prop}
\hat U_+^\mathrm{UDD,ideal} -\hat U_-^\mathrm{UDD,ideal} 
= {\cal O}((\gamma T)^{N+1}).
\end{equation}
Combined with \eqref{eq:add1} this equation implies \eqref{eq:bound1}.

The bound $(\gamma T)^{N+1}$ resulting from the sequence can be 
improved systematically by enlarging $N$. 
The bound $N(\gamma \tau_\mathrm{mx})^2$ resulting from the pulses can be 
improved by making it shorter. But if this is not possible,
one is stuck because the exponent of 2 cannot be incremented
for $\theta=\pi$ pulses as implied by mathematical no-go theorems
\cite{pasin08a,pasin08b}.
Hence we are facing here a serious conceptual obstacle.

Recently a variant of \eqref{eq:ansatz-dd}
\begin{subequations}
\label{eq:ansatz-av}
\begin{eqnarray}
\label{eq:ansatz2}
 \hat U_\mathrm{p}(t+\tau_\mathrm{p},t) &=&
\hat  U_\mathrm{p}^\mathrm{zero}(t+\tau_\mathrm{p},t)  + 
{\cal O}((\gamma\tau_\mathrm{p})^M)\qquad
\\
\label{eq:zero}
\hat U_\mathrm{p}^\mathrm{zero}(t+\tau_\mathrm{p},t) &=& 
e^{-\mathrm{i}\tau_\mathrm{p} \hat A_0}
\hat{P}_\theta 
\end{eqnarray}
\end{subequations}
with $M=3$ was shown \cite{pasin09a} 
to reduce the correction to $(\gamma\tau_\mathrm{p})^3$.
 Note that in \eqref{eq:ansatz-av}
only the Hamiltonian $\hat A_0$ of the bath occurs without
coupling to the spin. Hence $[\hat A_0,\hat{P}_\theta]=0$ holds
and no $\tau_\mathrm{s}$ needs to be introduced. 

Explicit solutions are obtained for pure dephasing \cite{pasin09a}.
The correlation time of the dephasing bath should not
be much smaller than $\tau_\mathrm{p}$. 
Moreover, no no-go theorem was found which prevents
to achieve higher orders as well.
Indeed, a recursive scheme based on concatenation is proposed
which achieves arbitrary order $M$ at exponential cost \cite{khodj09c}, i.e.,
each composite $\pi$ pulse consists of  $> 17^{M-1}$ elementary pulses.  
This demonstrates that in principle arbitrary $M$ can be achieved
though the exponential cost may spoil its practical usefulness.
But due to the shortness of the pulses compared to the whole sequence 
($\tau_\mathrm{p}\ll T$) we do not expect that particularly large values
of $M$ are required.

The property \eqref{eq:ansatz-av} is promising, but it \emph{cannot}
be used in standard DD,  or in  UDD in  particular,
as ersatz for an instantaneous pulse. 
This is so because any standard DD sequence
presupposes that between the pulses $\hat{P}_\pi$ the full Hamiltonian
$H$, not only $\hat A_0$, is active. This conceptual obstacle cannot be 
solved by pulse shaping because the no-go theorems block further 
progress\cite{pasin08a,pasin08b}. 
To overcome this obstacle is the main achievement of the present paper.
We find that an adjustment of the sequence to the pulses
of finite duration is required.

Our present fundamental observation is that relation \eqref{eq:zero} 
translates to $\widetilde F(t)$ for a single realistic pulse
between $t^-$ and $t^+=t^- +\tau_\mathrm{p}$ in the form
\begin{equation}
\label{eq:pulse-F}
\widetilde F(t) = \left\{
\begin{array}{ccc}
1 & \mathrm{for} & t<t^-\\
0 & \mathrm{for} & t^-<t<t^+\\
-1 & \mathrm{for} &  t>t^+
\end{array}
\right. .
\end{equation}
The $\pi$ pulse implies the inversion of the sign. But during the pulse
itself the relation \eqref{eq:ansatz-av} implies that the coupling between
spin (qubit) and bath is effectively averaged to zero
up to ${\cal O}((\gamma\tau_\mathrm{p})^M)$.
This is so since $\hat A_0$ in  \eqref{eq:zero} does not comprise
 the spin-bath coupling; it only comprises the bath dynamics.
This implies that the switching function $\widetilde F(t)$
takes the value zero during the pulse.
Note that there are jumps in the switching function even though the pulse 
is generated by  bounded control. The reason for this behavior is
that unitary time evolution is considered over \emph{finite} time intervals,
not over infinitesimal intervals.
This means that from the hierarchical level of the sequence
we do not look into the pulses. The description with $\widetilde F(t)$
is only valid on the level of the sequence, not within the pulse interval.
Furthermore, the correction term in Eq.\ \eqref{eq:ansatz2} may not be 
forgotten.

Next we look for a sequence with 
$\widetilde F(t)\in\{-1,0,1\}$ which corresponds to an
odd function $F(\vartheta)\in\{-1,0,1\}$ with the antiperiodic behavior
$F(\vartheta+\pi/(N+1))=-F(\vartheta)$. Such a sequence,
RUDD (realistic UDD), allows for the
same mathematical argument as UDD \eqref{eq:switch-theta-UDD}
ensuring that the effective time evolution of the spin is
the identity up to corrections of the order $(\gamma T)^{N+1}$. 
The reason is  the antiperiodicity of the switching function which is the 
fundamental reason for the annihilation of the preceding orders
\cite{yang08,pasin09c}. Note that this argument holds only
for UDD and similar optimized sequences. 
Hence we do not provide a general
scheme for the incorporation of pulses of finite duration
into arbitrary sequences.

\begin{figure}[ht]
    \begin{center}
    \includegraphics[width=0.99\columnwidth,clip]{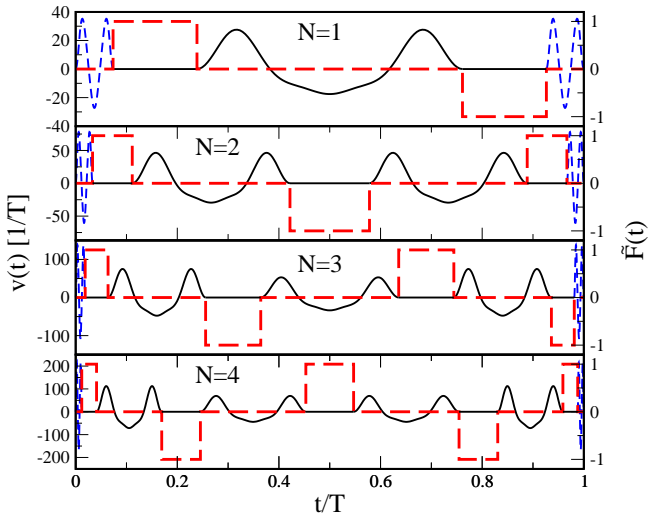}
    \end{center}
    \caption{Amplitudes $v(t)$ (black solid and blue short-dashed
      lines) for pulses rotating about 
      axes in the $xy$ plane and the resulting switching functions 
      $\widetilde F(t)$ (red dashed lines) for $N=1, 2,3,4$ $\pi$ pulses.
      For clarity, fairly large values $\vartheta_\mathrm{p}=0.7 \pi/(2N+2)$ 
      are chosen for the sake of clarity. For rendering purposes,
	$\vartheta_\mathrm{p}$ is chosen here to depend on $N$.
      Generally, it only has to fulfil $\vartheta_\mathrm{p} \le \pi/(2N+2)$.
      Black solid lines stand for $\theta=\pi$ pulses
      with $v_\pi(t)$ as in \eqref{eq:puls} with
      $a_\pi=10.804433[1/\tau_\mathrm{p}]$, 
      $b_\pi=6.831344[1/\tau_\mathrm{p}]$,
      $c_\pi=2.174538[1/\tau_\mathrm{p}]$. The first and the last
      pulse rotates by $2\pi$ (blue short-dashed lines)
      with $a_{2\pi}=10.236155[1/\tau_\mathrm{p}]$, 
      $b_{2\pi}=2.9661717[1/\tau_\mathrm{p}]$,
      $c_{2\pi}=0.889052[1/\tau_\mathrm{p}]$. For clarity, $v_{2\pi}(t)/5$
      is plotted.
      \label{fig:illustration}
}
\end{figure}

The sequence fulfilling the requirement of antiperiodicity reads
\begin{equation}
\label{eq:switch-theta-RUDD}
F_\mathrm{RUDD}(\vartheta) = \left\{ 
\begin{array}{cl}
(-1)^j &  \mathrm{for} \ \vartheta \in
\left(\frac{j\pi}{N+1}+\vartheta_\mathrm{p},
\frac{(j+1)\pi}{N+1}-\vartheta_\mathrm{p} \right)\\
0 &  \mathrm{otherwise}
\end{array} \right.
\end{equation}
for $j\in\mathbbm{Z}$. The Fourier series comprises
only the odd $\sin$ harmonics $\sin(l(N+1)\vartheta)$
with coefficients \mbox{$4\cos(l(N+1)\vartheta_\mathrm{p})/(\pi l)$}.
The parameter $0\le \vartheta_\mathrm{p} \le \pi/(2N+2)$ determines the
duration of the pulses. Except for the given inequality it is
independent of $N$. Note that
pulses of equal duration in $\vartheta$ do not correspond to
pulses of equal duration in time $t$
\begin{equation}
\label{eq:switch-t-RUDD}
\widetilde F_\mathrm{RUDD}(t) = \left\{ 
\begin{array}{cl}
(-1)^j &  \mathrm{for} \ t \in
\left(t_j^+,t_{j+1}^-\right)
\\
0 &  \mathrm{otherwise}
\end{array} \right.
\end{equation}
with $t_j^\pm:= T\sin^2\left[\frac{j\pi}{2N+2}
\pm\vartheta_\mathrm{p}/2\right]$.
This is illustrated for $N=\{1,2,3,4\}$ in Fig.\ 
\ref{fig:illustration} where also the necessary time-dependent amplitudes 
$v(t)$ defining the control Hamiltonian $H_\mathrm{C}(t)=v(t)\sigma_y$ 
are shown; $\sigma_y$ is the $y$ component of the Pauli matrices.
For instances, the amplitudes can be parametrized by
\begin{eqnarray}
 v_\theta(t)\!\!&=&\!\!
{\theta}/{2}+(a_\theta-\theta/2)\cos(2\pi t/\tau_\mathrm{p})+\nonumber\\
&& (b_\theta-a_\theta) \cos(4\pi t/\tau_\mathrm{p})+\nonumber\\
&& (c_\theta-b_\theta) \cos(6\pi t/\tau_\mathrm{p}) 
- c_\theta \cos(8\pi t/\tau_\mathrm{p}).\qquad
\label{eq:puls}
\end{eqnarray}

There is one subtlety about the beginning and the end
of the sequence. In order to generate the switching function
\eqref{eq:switch-t-RUDD}  there must be  a first and a last 
pulse which averages the coupling between 
spin and bath to zero while inducing \emph{no} net rotation.
For this purpose $\theta$ can take any multiple of $2\pi$. Solving
the equations derived in Ref.\ \onlinecite{pasin09a}, which 
imply that the pulse fulfills the relation
\eqref{eq:ansatz-av}, leads to the parameters
given in the caption of Fig.\ \ref{fig:illustration}.

The duration of the pulses is shortest towards the ends
of the interval $T$ for which the quantum state of the spin is to be stored.
Concomitantly, the amplitudes are largest for the first and the
last pulse. In practice, the initial and the final pulse can 
be combined with the pulses by which the quantum state of the
spin (the qubit) is generated, for instance a $\pi/2$ pulse
(for solutions see Ref.\ \onlinecite{pasin09a}).

What has been achieved by the sequence \eqref{eq:switch-t-RUDD}
depicted in Fig.\ \ref{fig:illustration}? This sequence is an
optimized dynamic decoupling scheme made from pulses 
of finite duration and finite amplitudes with analytically
founded properties. Bounded control is a crucial aspect
for realistic sequences, so that the proposed sequence
is an important step closer to a realistic scenario.
Nevertheless, the sequence is still
optimized in the sense that it shares the same power law
property as the UDD built from instantaneous pulses.

There are two sources for corrections in the unitary
time evolution  $\hat U_\pm^\mathrm{RUDD}$ of  the RUDD sequence. The first
kind of corrections stems from the pulses which
are only close to $\hat  U_\mathrm{p}^\mathrm{zero}(t+\tau_\mathrm{p},t)$
but not identical to it, see Eq.\ \eqref{eq:ansatz-av}.
Denoting the time evolution of the RUDD sequence made 
from the pulses $\hat  U_\mathrm{p}^\mathrm{zero}(t+\tau_\mathrm{p},t)$
in \eqref{eq:zero} by $\hat U_\pm^\mathrm{RUDD,zero}$ we know
from \eqref{eq:ansatz2}
\begin{equation}
\label{eq:add2}
\hat U_\pm^\mathrm{RUDD} = \hat U_\pm^\mathrm{RUDD,zero} +
{\cal O}(N(\gamma\tau_\mathrm{mx})^M)
\end{equation}
for $N$ pulses. The second kind of corrections stems
from the sequence itself.
The time evolution $\hat U_\pm^\mathrm{RUDD,zero}$ is
rigorously governed by $F_\mathrm{RUDD}(\vartheta)$ 
defined in \eqref{eq:switch-theta-RUDD}. Then we know
from Ref.\ \onlinecite{yang08} that
\begin{equation}
\label{eq:rudd-prop}
\hat U_+^\mathrm{RUDD,zero} -\hat U_-^\mathrm{RUDD,zero} 
= {\cal O}((\gamma T)^{N+1}).
\end{equation}
The total correction is given by the sum of both kinds
of corrections because their norm is invariant under
unitary transformations. So in analogy to \eqref{eq:bound1} we obtain
\begin{equation}
\label{eq:bound2}
\hat U_+^\mathrm{RUDD} - \hat U_-^\mathrm{RUDD} =
{\cal O}\left((\gamma T)^{N+1}\right) + {\cal O}
\left(N(\gamma \tau_\mathrm{mx})^M\right).
\end{equation}
We stress that this relation excludes 
mixed terms $(\gamma T)^{n}(\gamma\tau_\mathrm{mx})^m$ with $n\le N+1$
and $m<M$ because each pulse complies with \eqref{eq:ansatz2} separately.
We point out that $M$ does not need to be as large as $N$ because
the pulses are  much shorter anyway. So the relatively
short and simple pulse found in Ref.\ \onlinecite{pasin09a} realizing
$M=3$ may often be completely sufficient.

For later reference, we point out that the above derivation
also holds if we allow for an explicit analytic time dependence of the
operators $\hat A_0$ and $\hat A_1$ in \eqref{eq:hamilton}. This was recently
shown by us in the context of optimized dynamic decoupling
for time dependent Hamiltonians\cite{pasin09c} relying only on the mathematical
properties of the switching function $F(\vartheta)$.
Hence the same argument also applies to the RUDD sequence
if the pulses are shaped to realize zero coupling during
their duration, see Eq.\  \eqref{eq:pulse-F}. This is
definitely the case if there is no time dependence \emph{during}
the pulses because the pulses suggested in Refs.\ 
\onlinecite{pasin09a} and \onlinecite{khodj09c}
can be used. This is indeed a relevant case as we will discuss below.

We emphasize that the RUDD with \eqref{eq:bound2} 
provides an efficient scheme for dynamic decoupling based on bounded control.
It is the main result of our paper.
The previously obstructive no-go theorems \cite{pasin08a,pasin08b}
can be circumvented by the RUDD approach.
The qualitative novel finding in the present
work is that the sequence has to be adjusted
in a precise way in order to allow for realistic pulses 
while preserving the properties of the sequence of ideal pulses. 
Above we constructed a precise prescription which achieves
the necessary adjustment. We expect
that this observation extends beyond the case of UDD and RUDD.
This expectation is illustrated in the next section.

{We emphasize that the number of pulses $N$ cannot be
made infinite without using shorter and shorter pulses with
larger and larger amplitudes. Hence a given bound to the available
power of the control pulses limits the maximum possible number of pulses
for a given interval $T$. But such limits exist
in any experimental setup anyway \cite{bierc09a,bierc09b,du09} and we
expect that the RUDD approach will prove its usefulness for
a moderate number of pulses. The limit $T\to 0$ is studied here
to characterize the mathematical properties of the idealized
situation. The achievement of the RUDD over the UDD sequence is
that for any finite duration $T$ and finite number of pulses $N$
only pulses of finite amplitude are needed.}

\section{\textbf{Iterated} Sequences}

In view of the above we expect that the famous CPMG 
sequence \cite{carr54,meibo58} can be improved
for realistic pulses as follows. The CPMG  is given by
the $n$-fold iteration of the two-pulse cycle
$t-\pi-2t-\pi-t$, where $\pi$ stands for a $\pi$ pulse and $t$ for
free evolution of time $t$. This two-pulse cycle is
the UDD sequence for $N=2$ pulses \cite{uhrig07}. Hence for pulses of 
finite duration the iteration of the $N=2$ panel in Fig.\ 
\ref{fig:illustration} suggests itself. A slight modification
is possible by replacing two $2\pi$ pulses, where two cycles meet,
by one $2\pi$ pulse of double the length. Hence it is promising to
use the sequence 
\begin{eqnarray}
&&(2\pi)_{t_1}\left[-t_2-\pi_{\tau_\mathrm{p}}-2t_2-\pi_{\tau_\mathrm{p}}-t_2-
(2\pi)_{2t_1} \right]^{n-1}-
\nonumber \\
&& \qquad -t_2-\pi_{\tau_\mathrm{p}}-2t_2-\pi_{\tau_\mathrm{p}}-
t_2-(2\pi)_{t_1},
\end{eqnarray}
with $t_1=2t(1-\cos(\vartheta_\mathrm{p}))$, 
$t_2=2t\sin((\pi/6)-\vartheta_\mathrm{p})$,
and $\tau_\mathrm{p}=4t\cos(\pi/6)\sin(\vartheta_\mathrm{p})$. 
The subscripts indicate the pulse durations.
We iterate that the advocated recipe to account for bounded
control only applies to UDD-type sequences.

\section{Simulation of a RUDD sequence}

The advocated RUDD sequence relies on its mathematical properties
which have a certain beauty in themselves. 
But the ultimate check will be its experimental usefulness. A crucial step
on this route is an experiment with simulated noise
such as the one performed for  UDD \cite{bierc09a,bierc09b,uys09b}.
There, the simulated noise was switched off
during the pulse. The theoretical calculations took this dead time
of the noise into account. But variable pulse lengths
such as in  RUDD were not considered.

We propose to implement the RUDD according to \eqref{eq:switch-t-RUDD}
with  pulses of finite, constant amplitudes
during the intervals where $\widetilde F(t)=0$. No pulse
shaping is required if the noise is switched off during the pulse
so that $\widetilde F(t)=0$ is fulfilled by construction. 
Hence we have
\begin{equation}
\label{eq:bound-simulate}
\hat U_+^\mathrm{RUDD} - \hat U_-^\mathrm{RUDD} =
{\cal O}\left((\gamma T)^{N+1}\right) 
\end{equation}
for this particular experiment instead of \eqref{eq:bound2}.

The pulse intervals have to be chosen as in \eqref{eq:switch-t-RUDD}. 
Concomitantly the amplitudes have to vary
to ensure that the pulses are $\pi$ pulses. In this way any
deviation resulting from the pulses is eliminated. It is highly
interesting to investigate if such a  RUDD sequence is more
powerful than existing realizations.

\section{Longitudinal relaxation}

A UDD sequence 
can also suppress longitudinal relaxation \cite{yang08}. Pulses of angle
$\pi$ about the $z$ axis can suppress  terms
proportional to $x$ and $y$ component, i.e., $\sigma_x$ and $\sigma_y$,
of the Pauli matrices  up to order $(\gamma T)^{N+1}$
for $\{t_j\}$ as in \eqref{eq:switch-t-UDD}. Concatenation of 
such UDD sequences  (CUDD) can be used to suppress any kind of relaxation 
\cite{uhrig09b}. The QDD appears to be the most efficient scheme
to fulfill this purpose \cite{west09,pasin09c}.

The pulses depicted in Fig.\ \ref{fig:illustration}
and computed in Ref.\ \onlinecite{pasin09a}
also work to order $(\gamma \tau_\mathrm{p})^3$ if used for rotations
$\hat P^z_\theta$ around the $z$ axis for arbitrary couplings to
$\sigma_x, \sigma_y$, and $\sigma_z$. 
The pulse $\hat P^z_\pi$ induces an inversion of
the sign of the couplings along  $\sigma_x$ and $\sigma_y$. 

To see this one has to modify the specific calculation for a
rotation about a fixed axis in Ref.\ \onlinecite{pasin09a}
according to $\hat A_0 \to \hat B_0=
\hat A_0 + \hat A_z \sigma_z$ and
$\vec{\sigma}\cdot\vec{A} \to \vec{\sigma}_\perp\cdot\vec{A}=
\sigma_x \hat A_x +  \sigma_y \hat A_y$ 
for a Hamiltonian  $H=\hat A_0 +\vec{\sigma}\cdot\vec{A}$.
The Eq.\ (\ref{eq:ansatz-av}) becomes
\begin{equation}
 \label{eq:ansatz-lr}
\hat U_\mathrm{p}(t+\tau_\mathrm{p},t) =
e^{-\mathrm{i}\tau_\mathrm{p}\hat B_0}
\hat{P}^z_\theta {\hat U_G}(\tau_\mathrm{p},0),
\end{equation} 
where ${\hat U_G}(\tau_\mathrm{p},0)$ encodes
the corrections. It is given by
\begin{equation}
 {\hat U_G}(\tau_\mathrm{p},0)=
T\left\{ e^{-\mathrm{i}\int_{0}^{\tau_\mathrm{p}}  G(t) \mathrm{d}t}\right\}.
\label{U_G}
\end{equation}
where the time dependent Hamiltonian of the corrections
 $G(t)$ stands for
\begin{equation}
 \label{eq:G} 
G(t):=e^{\mathrm{i}\hat B_0 t} (\hat{P}^z_t)^\dagger
\left(\vec{\sigma}_\perp\cdot\vec{A}\right) \hat{P}^z_t 
e^{-\mathrm{i}\hat B_0t}
\end{equation} 
with $\hat{P}^z_t=\mathrm{exp}\left(-i\sigma_z
\int_0^t\mathrm{d}s\  v(s)\right)$ resulting from 
$H_\mathrm{C}(t)=v(t)\sigma_z$
representing the pure control rotation at instant $t$.
We show that ${\hat U_G}(\tau_\mathrm{p},0)= {1}
+{\cal O}((\gamma\tau_\mathrm{p})^3)$ if the shaped rotations about
$\sigma_y$ proposed in Ref.\ \onlinecite{pasin09a}
are applied about $\sigma_z$.

The Magnus expansion \cite{magnu54,haebe76}
 allows us to write the time evolution in Eq.\ (\ref{U_G}) 
in terms of  cumulants 
$U_G(\tau_\mathrm{p},0)=\mathrm{exp}\left(-i\tau_{\mathrm p}\sum_{i=1}^\infty 
\eta^{(i)}\right)$. 
Each cumulant $\eta^{(i)}$ scales as $(\gamma\tau_{\mathrm p})^i$. 
Following the approach of Ref.\ \onlinecite{pasin09a}, it is 
straightforward to find
\begin{equation}
 \label{eq:eta1} \eta^{(1)}=\eta_{11}\left(\sigma_x \hat A_y-\sigma_y 
\hat A_x\right)+\eta_{12}\left(\sigma_x \hat A_x+\sigma_y \hat A_y\right),
\end{equation}
with $\eta_{11}$ and $\eta_{12}$ the first order corrections. 
For the second order one finds
$\eta^{(2)}=\eta^{(2a)}+\eta^{(2b)}$ with
\begin{subequations}
\begin{eqnarray}
 \eta^{(2a)}&=&\eta_{21}[\hat B_0,
\sigma_x \hat A_y-\sigma_y \hat A_x]
\nonumber
\\
&& +\eta_{22}[\hat B_0,
\sigma_x \hat A_x+\sigma_y \hat A_y]\qquad
\\
 \eta^{(2b)} &=& 2\eta_{23}\sigma_z \left\{[\hat A_y,\hat A_x]+i(\hat A_x^2+
\hat A_y^2)\right\}. 
\end{eqnarray} 
\end{subequations}
The expressions for $\eta_{21}$, $\eta_{22}$, and $\eta_{23}$ are 
\begin{subequations}
\label{etas_specific}
\begin{eqnarray}
 \label{eta11_specialcase} 
\eta_{11} &:=& \int_0^{\tau_\mathrm{p}} \mathrm{d}t \sin\psi(t)
\\
 \label{eta12_specialcase} 
\eta_{12} &:=& \int_0^{\tau_\mathrm{p}}\mathrm{d}t  \cos\psi(t)
\\
 \label{eta21_specialcase} 
\eta_{21} &:=& \int_0^{\tau_\mathrm{p}}\mathrm{d}t\ t\ \sin\psi(t)
\\
 \label{eta22_specialcase} 
\eta_{22} &:=& \int_0^{\tau_\mathrm{p}}\mathrm{d}t\ t\ \cos\psi(t)
\\
 \label{eta23_specialcase} 
\eta_{23} &:=& \iint_0^{\tau_\mathrm{p}} \!\!\!
\mathrm{d}t_1\mathrm{d}t_2 \sin(\psi({t_1})-\psi({t_2}))\text{sgn}(t_1-t_2),
\qquad
\end{eqnarray} 
\end{subequations}
where $\psi(t):=2\int_0^t v(t')dt'$.
These conditions are exactly the same as 
those reported in Ref.\ \onlinecite{pasin09a} for pure dephasing.
Hence they have the same solutions and
the pulses depicted in Fig.\ 1 make the first and the second order 
corrections vanish also for longitudinal relaxation. No changes
in the pulse shapes are required.
Up to the third order, the transverse coupling is 
suppressed and only  the $z$-coupling survives unaltered. Between two 
subsequent pulses the sign of the $x$ and $y$ coupling is inverted.
For pulses corrected in higher order $M>3$ corrections we again refer to Ref.\ 
\onlinecite{khodj09c} for a proof-of-principle construction.
Hence, the RUDD sequence is equally applicable for the suppression of
longitudinal relaxation.

{Eqs.\ (\ref{etas_specific}) hold generally for the suppression of 
decoherence perpendicular to the fixed axis of rotation of the pulse. 
The decohering coupling along this axis is not suppressed. The case
of pure dephasing can be seen as special case of  the more general
case discussed here: There is no coupling along the axis of rotation and
only one (out of two possible) perpendicular coupling.}

\section{Concatenation of RUDD Sequences}

To tackle general decoherence the combination of
at least two sequences of rotations about perpendicular
spin axes are used. Available schemes rely on recursive
concatenation as for CDD \cite{khodj05} or CUDD \cite{uhrig09b} or on
a single step concatenation as for QDD \cite{west09,pasin09c}. Hence it is
natural to consider concatenation of RUDD sequences of
rotations about two perpendicular spin axes.

For simplicity we consider
the QDD scheme which comprises two levels. On the first
level two (e.g., $\hat A_x$ and $\hat A_y$) 
of the three couplings $\vec{A}$ to the components of
$\vec{\sigma}$ are eliminated up to a certain order.
This is exactly what is achieved by a RUDD of $N_z$ rotations
about $z$ for longitudinal
relaxation as discussed in the previous section.\footnote{
We use the term `longitudinal relaxation' for the couplings
perpendicular to the $z$-coupling and eliminate them first. But we
stress that in our theoretical treatment no spin axis is
special. Thus the first suppression can also be done by $x$-
or $y$-rotations suppressing the corresponding perpendicular
couplings, i.e., the $y$ and $z$ couplings or the $x$ and $z$
couplings, respectively.}
Up to the corrections ${\cal O}\left((\gamma T_z)^{N_z+1}\right) + {\cal O}
\left(N(\gamma \tau_{z,\mathrm{mx}})^M\right)$  of the primary level, 
the resulting time evolution is given an effective Hamiltonian
which implies dephasing only. Note that $T_z$ is the duration of
the primary RUDD sequence. 

The effective Hamiltonian
is of the form  given in 
Eq.\  \eqref{eq:hamilton}, but with time dependent operators
$\hat A_0(t)$ and $\hat A_z(t)$. The time dependence of 
these operatores is analytical since it results from the
time evolution for the time interval $T_z$ 
given by the Schr\"odinger equation on the primary level.
Note that it is understood that all the switching
instants are chosen relative to $T_z$.
Then a UDD sequence of duration $T_\perp$ can be applied on the
secondary level\cite{pasin09c} which
consists of $N_\perp$ rotations about the spin $x$ or
$y$ axis to suppress dephasing up to corrections 
${\cal O}\left((\gamma T_\perp)^{N_\perp+1}\right)$.
This means that general decoherence can be suppressed
by a RUDD on the primary and a UDD on the secondary level.

To obtain a quadratic scheme using pulses of finite duration
we use a RUDD also on the secondary level, calling the resulting
scheme QRUDD. This is possible since in the derivation
of the RUDD for pure dephasing in Sect.\ \ref{sect:dephasing}
we mentioned that the initial Hamiltonian may display an analytic
time dependence. This effective time dependence results here from
the pulse sequence on the primary level. Thus it is by construction
not present \emph{during} the secondary pulses. These secondary pulses
have to be constructed in the presence of general decoherence
$\vec{\sigma}\cdot\vec{A}$ so that the explicit solution in
Ref.\ \onlinecite{pasin09a} cannot be used. But the equations
to be solved are given in sufficient generality in this reference.
For a proof-of-existence we refer to the work by Khodjasteh {\it et al.} 
where  concatenated solutions for such pulses are constructed recursively
\cite{khodj09c}.

Hence, the known mathematical properties of UDD sequences and
of $\pi$ pulses suffice to conclude that even general decoherence
can be efficiently suppressed by dynamic decoupling with bounded
control by means of this QRUDD scheme.
It is a quadratic scheme of  UDD sequences of bounded, and thus
essentially realistic, control pulses.

\section{Summary}

We derived in this paper that an optimized sequence of 
realistic pulses (RUDD), i.e., of finite duration and
amplitude, can be set up which suppresses dephasing or
longitudinal relaxation up to $T^{N+1}$ in the length of the
sequence and up to $\tau_\mathrm{mx}^M$ in the maximum duration of the pulses,
avoiding all mixed terms in contrast to previous proposals.

This statement is based on rigorous analytical calculations
for bounded baths and it
is expected to apply to systems with hard high-energy cutoff.
Our argument is based on the fundamental
mathematical property of the optimized sequences of UDD-type,
namely a certain antiperiodicity in the auxiliary variable $\vartheta$.
Thus it only applies to such sequences and  not to arbitrary sequences.

We introduced and exploited the concept of double scaling
in the durations $\tau_\mathrm{p}$ of the pulses
\emph{and} in the duration $T$ of the whole sequence.
We emphasize that both scales can be varied independently
except for a certain constraint, see Eq.\ \eqref{eq:constraint}.

The key achievement  is to establish a precise prescription
how the sequence has to be adjusted to allow for the use of
 pulses with bounded amplitudes, which are thus decisively
more realistic.
Only the adjustment of the sequence to the use
of tailored pulses of finite duration allowed us to 
circumvent no-go theorems 
\cite{pasin08a,pasin08b} concerning the properties of tailored pulses.

The proposed RUDD can be used for suppressing pure dephasing, i.e.,
suppressing coupling of the bath to one spin component,
or for suppressing longitudinal relaxation, i.e., 
suppressing coupling of the bath to two spin components.
General decoherence, i.e., 
suppressing coupling of the bath to all three spin components,
 cannot be suppressed by a single RUDD but by a quadratic
concatenated scheme (QRUDD) of two RUDDs, made from rotations
about two perpendicular spin axes.

Based on the known properties of 
UDD \cite{uhrig07,lee08a,yang08} and QDD
\cite{west09,pasin09c}, we think that the design of the sequences is
very close to its optimum. But we expect that the design of
the pulses can still be improved. While for $M=3$ (leading non-vanishing
correction is cubic in $\tau_\mathrm{p}$) rather simple pulse shapes
are known \cite{pasin09a}, for higher order pulses with $M>3$ recursive
concatenation provides a recipe for their construction at the expense
of an exponential increase in the number of elementary pulses \cite{khodj09c}.

Certainly, further research is called for to determine the performance of
RUDD and QRUDD for specific models. One important issue
is to determine the size of the prefactors of the neglected terms.
Another issue on the way to the experimental application 
of RUDD and QRUDD is to investigate the robustness of both
the tailored pulses and the sequences to imperfections such as
imprecise timing.

To stimulate further research on the experimental side we
proposed an experimental setup to verify the RUDD for
simulated noise which can be switched off \cite{bierc09a,bierc09b,uys09b}
so that a RUDD can be checked without pulse shaping.

\begin{acknowledgments}
The financial support by the grant  UH 90/5-1
of the DFG is gratefully acknowledged.
\end{acknowledgments}

%\bibliographystyle{apsrev}
%\bibliography{../../bibinput/liter10}

\end{document}